\documentclass[
 reprint,
 amsmath,amssymb,
 aps,
 prl,
 superscriptaddress,
 groupedaddress,
 floatfix,nofootinbib
]{revtex4}

\usepackage{graphicx}
\usepackage{dcolumn}
\usepackage{bm}
\usepackage{hyperref}

\begin{document}

\preprint{APS/123-QED}

\title{The importance of charged particle reactions in the r-process \\ on supernovae and neutron stars
}

\author{P.V.~Guillaumon}
    \email{guillaumon@if.usp.br}
\author{I.D.~Goldman}
 \affiliation{Universidade de Sao Paulo Instituto de Fisica \\ Rua do Matão 1371, 05508-090 Sao Paulo, Brazil}
 
\date{\today}

\begin{abstract}

We propose a $(p,xn)$ mechanism with dynamic production as a new set of nuclear reactions that could produce high density neutrons and explain the r- and rp-elements. We calculate the rate of thorium and uranium produced by our proposed mechanism and show that it is compatible with different stellar conditions found in explosive events at an initial temperature of $T \geq 3\times 10^{9} K$ with a  ``freeze-out'' by a neutrino-driven wind. We show that charged particle reactions could explain the  discrepancies in the abundances of ${}^{232}Th$ and ${}^{235,238}U$ nucleochronometers. We extend the endpoint of the rapid proton (rp) process far beyond the previous work by showing that $(p,xn)$ reactions could contribute to the nucleosynthesis of heavy stable neutron deficient nuclides, like ${}^{190}Pt$, ${}^{184}Os$, ${}^{180}W$ and ${}^{174}$Hf. This implies in a broader definition of the rp-process and has important consequences for the nucleosynthesis of heavy elements. We show that we did not need to assume an extreme condition for the drip line of super neutron-rich nuclei.

\end{abstract}

\maketitle




There has been an extensive debate about the origin of the neutrons for the r-process and its  astrophysical site. Could the r- and rp-processes be unified in neutron-star mergers and produce high density neutron-flux? As  \citet{Cameron2001} pointed out, many of the hypothesis about the r-process are ``an article of faith among astrophysicists", since ``the r-process operates in a much more complicated way than traditionally thought". Although the main mechanism of it is understood, i.e., high density neutrons captures ($\sim10^{20} \, s^{-1} \cdot cm^{-3}$) by neutron-rich nuclei up to the drip line, and  successive $\beta$-decay to the stability valley \cite{B2FH}, the astrophysical conditions and nuclear reactions remains unknown. Several possible nuclear reactions have already been proposed with more or less success to explain the r-elements of the Solar System, \cite{Cowan1991-267}. All those nuclear reactions share certain aspects in common: charged-particle reaction in light nuclei producing neutrons, necessity of explosive events for those reactions take place and optimistic conditions for the neutron binding energy in super neutron-rich nuclei.  


Astrophysical models based on supernova explosions and jets, and neutron-star-neutron-star collisions seem the most plausible, based on the fact that Solar System r-elements abundances (i.e, abundances due r-process) are reproducible with better accuracy by those models, \cite{Cowan1991-267,Cameron2003,Cameron2001,Sumiyoshi2001,Wanajo2001,Nishimura2015,Chornock2017}. This rely on the abundances of protons and electrons during all the evolution of stars, so $p(e^-,\nu_e)n$ can take place as a neutron source, despite the cross-section of $10^{-47} \, cm^{2}$, \cite{Wallerstein}.
More recently, LIGO collaboration identified a signature of actinides being produced by neutron-stars collisions, providing strong evidence for this site as a candidate for the r-process, \cite{Pian2017,Bartos2019}.

Nucleosynthesis models depend on the knowledge of nuclear masses, half-lives and neutron capture's cross sections of super-neutron rich nuclei whose experimental values are almost unknown. Despite the fact that theoretical predictions have been proved correct for several nuclei involved in the r-process, cumulative errors in some of them can have a huge influence in the final abundances of  r-elements.
For example, \citet{MOLLER1997}, the main reference for r-elements data, overestimates many, like in the case of  ${}^{139}Sb$, which is statically incompatible with experimental results, \cite{Arndt2011}. Also, if we analyze our knowledge about neutron-rich isotopes, they are still far below the necessary for the typical r-path, like for the N=126 waiting point that have a gap of $\sim 40$ neutrons between the r-process path and our knowledge of the nuclear chart. If the half-lives are bigger than $1 ms$ with reasonable neutron capture cross sections, as expected by \citet{MOLLER1997}, it should not be so difficult to produce such isotopes in laboratory.
Recently, the large neutron capture cross-section of ${}^{88}Zr$ was measured, \cite{Shusterman2019}.
When compared with the expected theoretical value of $10 \, b$, the result of $8.61 (69)\times 10^5 \, b$ is incredible higher.  It is surprising that all the previous nuclides with cross-section above $10^5 \, b$ are all odd. A careful study of neutron-cross sections and other nuclear parameters for short half-lives isotopes are of special interest for nucleosynthesis.
These nuclides are produced in supernovae or neutron star mergers and can increase the conditions for the r-/rp-processes or even be a neutron poison. 

\begin{figure}[h]
 \centering
   \includegraphics[width=\columnwidth]{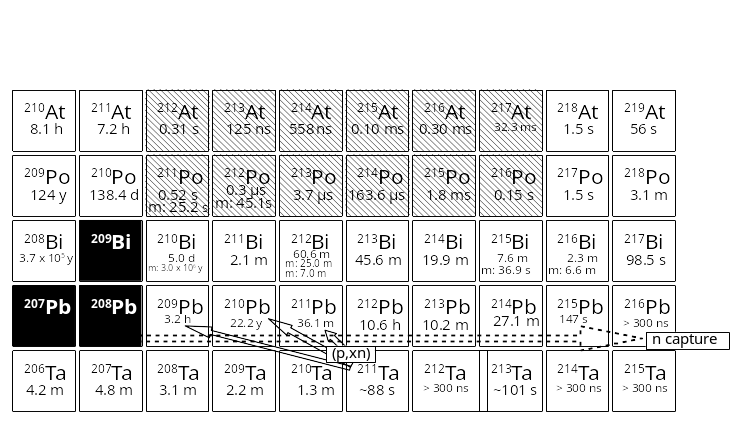}
 \caption{Chart of nuclides from Ta to At, with correspondent half-lives. Black squares refer to stable isotopes. The hatch ones, where the ``astrophysical structural barrier'' operates and they are prohibited for the r-path. We show a diagram of the $(p,xn)$ and $(n,\gamma)$ reactions, proposed is this paper as an unification of the r/i/rp-processes, and, for this reason, we call the $r^2p-$process (rrp-process).}
 \label{fig:fig1}
\end{figure}

\begin{figure}[h]
 \centering
   \includegraphics[width=\columnwidth]{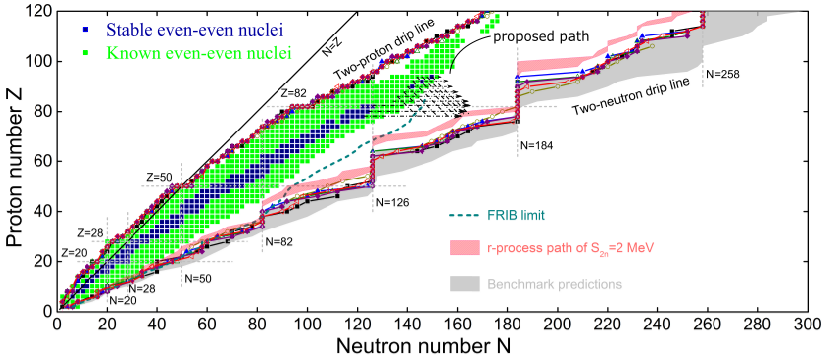}
 \caption{Chart of nuclides comparing the main paths for the r-process with the one we propose. Neutron drip line calculated by different mass formulas using density functional theory with four Skyrme interactions, relativistic interaction, and Weizsacker-Skyrme. Figure extracted and adapted from  \cite{Wang2015}. The common r-path populate isotopes from red to gray line, depending on mass formula and isotope network used, \cite{Arnould2007,Cowan1991-267,Wanajo2007,Thielemann2011,Kratz1988,Kratz1988_2,Mathews1990}.
 }
 \label{fig:fig8}
\end{figure}

We propose in this paper a different neutron source that could reproduce the r-elements with many advantages. Our model is based on a dynamic r-process whose neutrons come from $(p,xn)$ reactions on heavy elements while n-capture occurs. This kind of nuclear reactions has never been considered neither for the r- nor for the rp-process. Neutrino-driven winds in supernovae and neutron stars mergers would allow high energy protons ($\lesssim 50 \, MeV$)  to be mixed up with cold matter due a ``freeze-out'', \cite{Bliss2017,Perego_2014,Arcones2011,Arcones_2007}. In Fig. \ref{fig:fig1} it is possible to observe a schematic diagram of the proposed path. Note that above polonium, for neutron number $ 127 \leq N \leq 132$ and $Z \geq 84 $, the half-lives $T_{1/2} \lesssim \mu s $ makes this region prohibited for the r-process path, what is commonly called the ``astrophysical structural barrier''. Also, we should note that the neutron-rich isotopes of Ta, Pb and Bi have an increase of half-lives due their nuclear structure. With exception of ${}^{210m}Bi$ ($T_{1/2} = 3.0 \times 10^6 y$) and ${}^{209}Bi$ ($T_{1/2} = 2.1 \times 10^{19} y$), half-lives from minutes to years are pretty common. If we follow the r-path that lead us to the production of uranium and thorium from ${}^{208}Pb$ and ${}^{209}Bi$, we will notice that the whole path has this behavior. Comparing this r-path to the typical one, we need fewer assumptions about nuclear masses and half-lives of neutron-rich isotopes, since the majority of them has already been observed experimentally and can be used in astrophysical models. In Fig. \ref{fig:fig8} we compare the path proposed in this paper to produce uranium and thorium with the typical ones by several r-process calculations. It is possible to observe that our r-path goes a little beyond already known nuclei and closer to stability line, while the others are far away the known nuclei. 

Our model operates in a similar astrophysical condition of the $\alpha$-process proposed by \citet{Woosley1992}, where material in nuclear statistical equilibrium (NSE) at high temperature and with a large fraction of $\alpha$-particles are ``freeze-out''. Such a state occurs if a region of the star exceeds $ 5 \times 10^9 \, K $. At this condition we have a non-negligible ratio of particles with Gamow energy of $\sim 10 \, MeV$. If this $\alpha$-particles are mixed to the iron group nuclei and are expanded quickly enough, part of the $\alpha$-particles will have enough energy to surpass the Coulomb barrier of heavy elements. 
Depending on the initial conditions of neutron excess, $\alpha$  abundances and nuclear network employed, nuclei up to $A \sim 130$ can be produced, \cite{Howard1993,Bliss2017,Woosley1994,Farouqi2010}. Supposing an  artificially high-entropy wind, elements up to lead could be produced too, \cite{Farouqi2010,Freiburghaus1999,Meyer1992,Arcones2011}. 

These models based on $\alpha$-rich ``freeze-out'' still cannot explain sufficiently well the p-abundances of elements with $A \gtrsim 110$. Even with that consideration, \citet{Woosley1992} stated ``that the $\alpha$-rich freeze-out merges smoothly into an r-process, indeed, we believe, \textit{ the r-process}''. We would agree with them if elements like lead and thorium could be also explained. In all stars' evolution, protons remain the main particle, although decreasing in abundance. So, if we have a $ proton/\alpha$-rich ``freeze-out'', $(p,xn)$ and $(p,\gamma)$ reactions will be important and can explain heavier p/r-elements up to thorium and uranium. We call this process based on $proton$-rich ``freeze-out'' as the $r^2p$-process and we propose that this is the possible missing link to understand the r- and rp-processes. So the nucleosynthesis of elements heavier than iron, $\mathcal{N}$, could be explained by the sum of these two and the s-process:

\begin{equation}
    \mathcal{N} = \mathcal{N}_\alpha + \mathcal{N}_{r2p} + \mathcal{N}_{s},
\end{equation}

\noindent where $\mathcal{N}_\alpha$ stands for the $\alpha$-process, $\mathcal{N}_{r2p}$, for the $r^2p$, and $\mathcal{N}_{s}$, for the $s$-process. In a first approximation, for $A \gtrsim 110$, only the $r^2p$-process operates together with the slow process.

The differential equation considered, linking all reactions that create or destroy a nucleus by $r^2p$-process, includes $(p,xn)$ reactions rates, $EC,\alpha,\beta^\pm$ decays, and neutron captures.

We should add the rate of photo-disintegration reactions, i.e., the $-n_\gamma \sigma_{\gamma,n}Y(Z,A)$ term, for a complete description of this path. Without loss of generality, we just need that $\sigma_n n_n > \sigma_\gamma n_\gamma$. The $(\gamma,n)$ barrier can always be surpassed considering higher neutron densities, \cite{Goriely2008}, since photodisintegration cross-sections vary in the range $0.5-1 \, b$ at the Giant Dipole Resonance (GDR), in general, \cite{Varlamov1999}. In this paper we aim to show the importance of charged particle reactions for the nucleosynthesis of transbismuth and p-elements and to the r-process, so this picture is not affected by neglecting this term, although for a more precise description we should consider it, neutron's and proton's energy distribution, and stellar hydrodynamics.

It is important to note that the cross-section for $(p,xn)$ reactions are of the order of barn at energies below $50 MeV$. Since we should consider $(p,n)$, $(p,2n)$, and $(p,3n)$ reactions, this implies that the cross-section per neutron is $\sim 5 b / \textit{neutron}$! It is a huge increment when we compare with typical cross-sections for neutron production.

In our model, we considered $(p,1-3n)$ reactions in the isotopes ${}^{201-210}Pb/Bi$, which represent the main ones for neutron production from lead region. Starting with a mass of natural lead, we considered all the possible neutron captures from lead to uranium for a mass range between $ 201 \leq A \leq 250$, in a network of $\sim  350$ isotopes. We used TALYS code \cite{Koning2012} to simulate neutron and proton cross-sections. Whenever possible, we used experimental half-lives from NNDC database, \cite{NNDC}. In other cases, we used the semi-empirical formula of \citet{Liu2011}, which reproduces the half-lives of super neutron-rich nuclei better than Möller \textit{et al}, \cite{MOLLER1997,Moller2019}. Brute-forcing the whole input space, we searched for conditions of lead's mass, protons' flux and energies, and time duration that could reproduce the required conditions for the r-process.

\begin{figure}[h]
 \centering
   \includegraphics[width=\columnwidth]{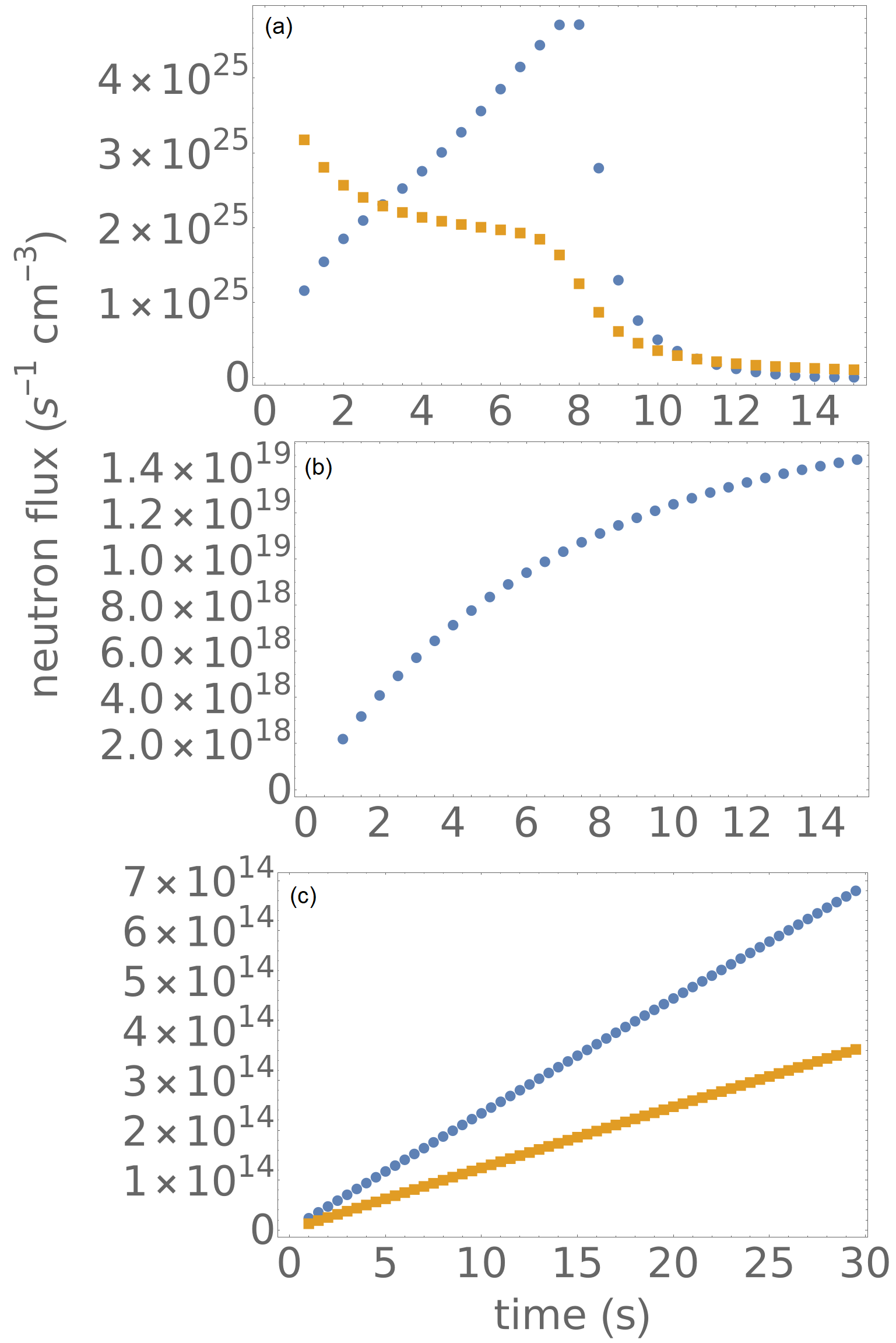}
 \caption{Neutron flux for the $r^2p$ mechanism developed in this work. (a) Lead shell of $N_{Pb} = 10^9 \, g$  , with protons of $E_p = 20 \, MeV$ and flux of $\phi_p = 10^{24} \, cm^{-3} \cdot s^{-1}$ ($\bullet$ blue dots). For the $\blacksquare$ orange dots, lead shell of $10^{12} \, g$  , with $E_pi_p = 18 \, MeV$ and $\phi_p = 10^{23} \, cm^{-3} \cdot s^{-1}$. In these conditions, it is possible to achieve $\sim 10^{25} cm^{-3} \cdot s^{-1}$.  We can observe a ``phase transition'' of the neutron-flux that in related to a steady-flow production. (b) $E_p = 26 \, MeV$  and $\phi_p = 10^{16} \, cm^{-3} \cdot s^{-1}$. (c) $N_{Pb} = 10^8 \, g$, $E_p = 26 \, MeV$ and $\phi_p = 10^{13} \, cm^{-3} \cdot s^{-1}$ ($\bullet$ blue dots); $N_{Pb} = 10^8 \, g$, $E_p = 30 \, MeV$ and $\phi_p = 10^{13} \, cm^{-3} \cdot s^{-1}$ ($\blacksquare$ orange dots). }
 \label{fig:neutron1}
\end{figure}

In Fig. \ref{fig:neutron1} (a) we note that our model is capable of producing $\sim10^{25} \textit{neutrons} \cdot s^{-1} \cdot cm^{-3}$ in less than 1 s. Our mechanism takes place for $20 \, MeV$ protons, with an initial mass of $10^9g$ ($\bullet$ blue dots) or $10^{12} g$ of lead ($\blacksquare$ orange dots), with a proton flux of $10^{24}$ or $10^{23} \textit{protons} \cdot s^{-1} \cdot cm^{-3}$, respectively. Considering that the Sun has $2\times 10^{30} kg$, where the mass fraction of hydrogen is $0.7491$, and the ratio hydrogen/lead in the Solar System is $10^{10}$ \cite{Lodders2003}, we have about $10^{20} kg$ of lead in a star analog to the Sun! Supernovae are, in general, much more massive than the Sun, so it is expected that the mass of lead is far beyond the one used in our model. A tiny admixture of the hydrogen envelope with heavy elements would be enough for the $r^2p$-process. The majority of low metallicity stars has some abundance of r-elements, providing one more evidence for the occurrence of the proposed process.

In Fig. \ref{fig:neutron1} (b), our model based on (p,xn) reactions reproduce the r-process, with 
$\sim 10^{19} \, \textit{neutrons} \cdot s^{-1} \cdot cm^{-3}$, protons of $26 \, MeV$, an initial mass of $10^{10} g $ lead in the shell that mixtures with the hydrogen envelope, and a proton flux of $10^{16} \, s^{-1} \cdot cm^{-3}$. As we shall see, a neutron flux of $10^{10-15} \, s^{-1} \cdot cm^{-3}$ is enough to produce uranium and thorium by the r-path suggested in this paper. This would be compatible with the i-process, \cite{Cowan-Rose1977}. In Fig.  \ref{fig:neutron1} (c), this condition is reproduced for protons of $26 \, MeV$ with $10^8 g$ of lead and $10^{13} \,  \textit{protons} \cdot s^{-1} \cdot cm^{-3}$. We prove that depending on the star's energy, proton's flux, and initial lead's mass, a steady-flow production may take place. This is a local equilibrium, so steady-flow models and NSE can not be used generally, despite the good agreement with the r-elements of the Sun. A better interpretation can be given as a phase transition of the neutron-flux in an explosive stellar event.



\begin{figure}[h]
 \centering
  \includegraphics[width=\columnwidth]{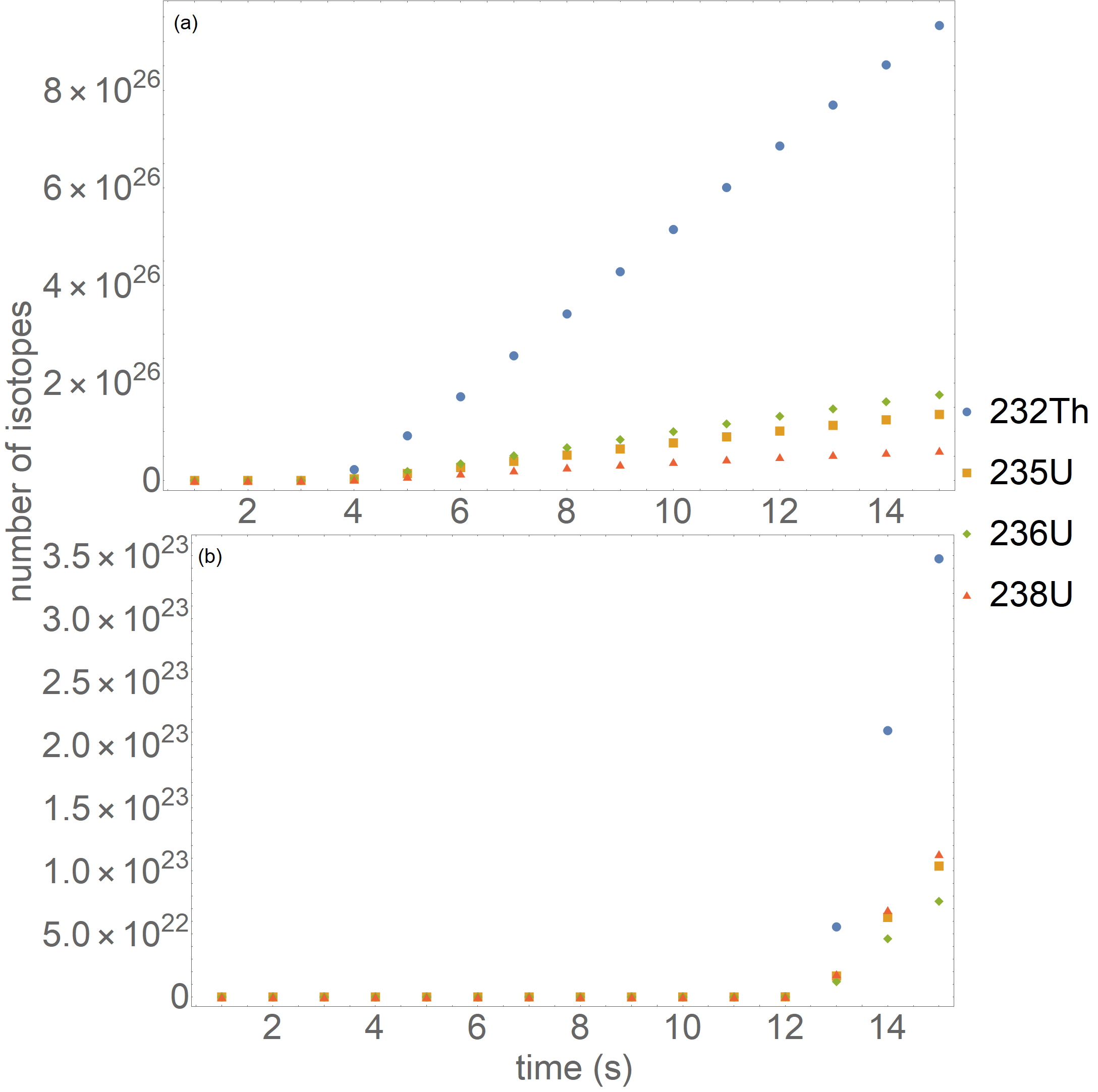}
 \caption{${}^{232}Th$ and ${}^{235,236,238}U$ abundances after short half-lives decayed. (a) Initial conditions: lead's shell of $N_{Pb} = 10^{9} \, g$, proton's flux and energy of $\phi_p = 10^{23} \, cm^{-3} \cdot s^{-1}$, and $E_p = 22 \, MeV$, respectively. Neutron energy of $E_n = 1.0 \, eV$. Only after $12 \, s$, uranium and thorium are exponentially produced. Since the half-lives for the r-path proposed are higher than the common one, this model can operate under low conditions, like the i-process. For the first seconds, although the neutron flux is not high enough for production of uranium and thorium, it can produce other r-elements. (b) $N_{Pb} = 10^{12} \, g$, $\phi_p = 10^{23} \, cm^{-3} \cdot s^{-1}$, $E_p = 19 \, MeV$, $E_n = 0.1 \, eV$. }
 \label{fig:fig10}
\end{figure}

We are interested in the total amount of ${}^{235,236,238}U$, and ${}^{232}Th$ produced by this mechanism, based in $(p,xn)$ reactions and dynamic production as a neutron source. If the proton's flux and energy are $\phi_p = 10^{28} \, cm^{-3} \cdot s^{-1}$ and $E_p = 30 \, MeV$, respectively, with lead's shell of $N_{Pb} = 10^{10} \, g$ and neutron's energy of $E_n = 0.1 \, eV$, the total amount of these nuclides reaches the secular equilibrium in less than one second.
Under less extreme astrophysical conditions, the time required for an explosive event to start the production of r-elements like uranium and thorium is greater than 10 s, as can be seen in Fig. \ref{fig:fig10} (a). In a mid-term condition, like that in Fig. \ref{fig:fig10} (b), after 5 s the secular equilibrium is reached. A more precise simulation should upgrade this nuclear reaction network, including an initial mass of other heavy elements with $A \geq 110$. Without loss of generality, it is trivial to note that the main results would be the same. This pattern, where neutron flux is increased until the critical value to continue the nucleosynthesis, is compatible with the hypothesis of hot bubbles and jets followed by a freeze-out, like in the $\alpha$-process. It will mainly depend on the duration of these jets or if they are made of successive flashes of matter or a big one. 

A consequence of  the $r^2p$-process is that we extend the endpoint of the rp process far beyond the previous work of  \citet{Schatz2001}. They calculated that it should end in the SnSbTe cycle, with similar astrophysical conditions of the $r^2p$-process. 
In Table \ref{tab:tab1} we calculated what should be the proton's flux to produce the heavy rp-elements from $(p,3n)$ reactions.
We assumed that target isotopes are produced just by other processes (or neglecting the contribution of the rp-process for their nucleosynthesis)  with the condition that the Solar System' abundance is reproduced. As we can see, we need about $10^{22-23} \, p \cdot s \cdot cm^{-3}$ to reproduce the Solar System' abundances. This is the same proton flux needed by the $(p,xn)$ reactions. We have done a crude estimate, without considering that isotopes with charge $Z-1$ will also contribute to the synthesis of this p-elements by proton reactions (like ${}^{121}Sb$ for ${}^{120}Te$). The half-lives of these natural isotopes, when not stable, are $> 10^{14} \, y$ and are negligible 
There are some discrepancies, mainly for low mass isotopes. A dynamic proton flux, with $\alpha$ and n reactions, photodisintegration, and a larger nuclear network could, in principle, explain them. Other astrophysical models that tried to explain nucleosynthesis of heavy rp-elements are mainly based on $(\gamma,n)$, $(\gamma,p)$ and $(\gamma,\alpha)$ reactions, \cite{Rapp_2006,Schatz1999,Rauscher_2002}, but there are still some discrepancies, mostly for the heaviest p-nuclei. Here we show that $(p,xn)$ reactions may solve these discrepancies, although never considered before in the nuclear networks of astrophysical models.

\begin{table}[]
\caption{Proton's flux required to produce Solar System's rp-abundances through (p,3n) reactions, assuming this process occurred in one second.}
\label{tab:tab1}
\begin{tabular}{|ccc|}
\hline \hline
Isotope      & Target                         & \shortstack{Flux \\ ($ 10^{23} \, p \cdot s \cdot cm^{-3}$)} \\ \hline
${}^{196}Hg$ & ${}^{198}Hg$ & 0.1-2.1                                     \\
${}^{190}Pt$ & ${}^{192}Pt$                      & 0.1-2.2                                     \\
${}^{184}Os$ & ${}^{186}Os$                      & 0.2-13                                      \\
${}^{180}W$  & ${}^{182}W$                       & 0.04-4.4                                    \\
${}^{174}Hf$ & ${}^{176}Hf$ & 0.3-93                                    \\
${}^{168}Yb$ & ${}^{170}Yb$                      & 0.4-2.3                                     \\
${}^{162}Er$ & ${}^{164}Er$                      & 0.9-65                                    \\
${}^{164}Er$ & ${}^{166}Er$                      & 0.6-1.3                                     \\
${}^{156}Dy$ & ${}^{158}Dy$                      & 6.1-20                                    \\
${}^{158}Dy$ & ${}^{160}Dy$                      & 0.4-1.4                                     \\
${}^{136}Ce$ & ${}^{138}Ce$                      & 9.3-670                                     \\
${}^{130}Ba$ & ${}^{132}Ba$                      & 16-980                                      \\
${}^{124}Xe$ & ${}^{126}Xe$                      & 0.5-3.4                                     \\
${}^{126}Xe$ & ${}^{128}Xe$                      & 22-290                                      \\
\hline\hline
\end{tabular}
\end{table}


The $\alpha$-process associated with the $r^2p$-process proposed in this paper could 
produce all the elements heavier than iron. This can occur in a high entropy neutrino wind provoking $\alpha$ and proton-rich freeze-out in neutron stars and supernovae. We believe that a complete r-elements production can be solved when all possible reactions for $r/i/rp/r^2p$-processes occur simultaneously and we will improve our code to account for them. 
The neutron-flux phase before the steady-flow, although do not produce enough r-elements, could produce lead from iron and compete with the $\alpha$-process. An adiabatic compression would work as a phase transition. 
The small amounts of lead and protons needed for this process, opens the possibility that $r^2p$-process occurred in second and third generation stars that do not produce them or in the Big Bang Nucleosynthesis (BBN).

We also note that other r-process models already proposed charged-particle jets with energies up to $200 \, MeV$ without even considering the importance of $(p,xn)$ and stripping reactions, \cite{Wanajo2003,Cameron2003,Wanajo2002,Cameron2001,Otsuki2000,Bethe1990,Woosley1994,Sale1986,Bonometto1985,Applegate1985}!

Since the proposed mechanism operates under extreme and intermediate neutron-flux to produce uranium and thorium, it may represent an unification for the rp/r/i-processes. The last one still need more experimental evidence, \cite{Denissenkov2017,Clarkson2017,Asplund1998}.

In conclusion, we demonstrated that the mechanism proposed in this paper is capable of producing r-elements in a different r-path than the typical ones. In association with the $\alpha$-process, the proposed mechanism is able to produce all the r-, s- and p-elements, and show the importance of charged particle reactions for the nucleosynthesis of the r-process. We have enough evidence to believe that a bound neutron in a super neutron-rich isotope is prohibited by nuclear structure and neutron bind energy, although more experimental evidence is still needed. We show that a steady-flow production is probably a local equilibrium in the neutron flux-phase space diagram. Depending on the stellar evolution in an explosive event, only this phase may be the main responsible of the r-process nucleosynthesis. Since this process operates under a rapid proton process, \cite{Wallace1981}, to generate the r-process by dynamic production, we present a possible unification for the r-  and rp-processes, and, for this reason, we call this the $r^2p-$process. Also, we extend the possible endpoint of the rp-process to include heavy elements like lead, bismuth and thallium, and above the SnSbTe cycle estimated by \citet{Schatz2001}. Since both processes are expected to occur in neutron-stars, \cite{vanWormer1994,Schatz1999,Pian2017}, it seems natural that they are interconnected . We show that $(p,xn)$ reactions may be important for both processes, although never considered before.

This paper is just a starting point into this area of study and seems to be a very promising solution to neutrino-driven nucleosynthesis in supernovae and neutron stars mergers. Without consideration of stellar hydrodynamics, it will not be possible to calculate accurately the solar abundances. Indeed, we propose a mechanism that could produce all the heavy elements and maybe solve the intriguing question about the site of the r-process.

\bibliography{bib}

\end{document}